\documentclass[12pt]{iopart}

\usepackage{bm}
\usepackage{amssymb}
\usepackage{graphicx}

\usepackage{bm}
 \usepackage{color}

\newcommand{\dd}{\text{d}}

\newcommand{\kk}{{\text{th}}}
\newcommand{\beq}{\begin{equation}}
\newcommand{\eeq}{\end{equation}}
\newcommand{\beqa}{\begin{eqnarray}}
\newcommand{\eeqa}{\end{eqnarray}}
\newcommand{\nn}{\nonumber\\}

\newcommand{\zetat}{\widetilde{\zeta}_{1}}

\newcommand{\text}[1]{\mathrm{#1}}

\eqnobysec

\begin{document}

\title[Impurity in a granular gas under Couette flow]{Computer simulations of an impurity in a granular gas under planar Couette flow}
\author{F Vega Reyes, A Santos and V Garz\'o}
\
\address{Departamento de F\'isica, Universidad de
Extremadura, E--06071 Badajoz, Spain}

\eads{\mailto{fvega@unex.es}, \mailto{andres@unex.es}, \mailto{vicenteg@unex.es}}

\date{\today}
\begin{abstract}
{We present in this work results from  numerical solutions, obtained by means of the direct simulation Monte Carlo (DSMC) method, of the Boltzmann and Boltzmann--Lorentz equations for an impurity immersed in a granular gas under planar Couette flow. The DSMC results are compared with the exact solution of a recent kinetic model for the same problem. The results  confirm that, in steady states and over a wide range of parameter values, the state of the impurity is enslaved to that of the host gas: it follows the same flow velocity profile, its concentration (relative to that of the granular gas) is constant in the bulk region, and the impurity/gas temperature ratio is also constant.
We determine also the rheological properties and nonlinear hydrodynamic transport coefficients for the impurity, finding a good semi-quantitative agreement between the DSMC results and the theoretical predictions.}
\end{abstract}
\noindent{\it Keywords\/}: granular matter, kinetic theory of gases and liquids,
rheology and transport properties

 \maketitle

\section{Introduction}\label{sec1}

The transport of granular matter has a growing interest for industrial and technological purposes. For this reason, it is convenient to study the behavior of simplified granular systems, both from an experimental and a theoretical point of view (see \cite{AT06}, for example, for a recent review on the field). Some of the phenomena of interest for industrial applications are the behavior of transport, diffusion, and segregation of grains depending on their different physical properties (mass, form, size, or inelasticity) \cite{KJN93,JY02,TAH03,K04,BRM05,BRM06,SGNT06,ATH06,G06a,MPEU07,G08,G09,SNTG09,GV10,K10}.
When the granular medium is dilute and vigorously shaken, the motion of grains resembles that of atoms or molecules in an ordinary gas and the near-instantaneous binary collisions prevail. Under these conditions, kinetic theory properly modified to account for the inelasticity of collisions provides a useful framework to analyze granular flows. It has been shown that  the corresponding kinetic equation describing the statistics of this many-particle system can generate the so-called `normal solution', in which all the spatial dependence occurs through the average fields \cite{BDKS98}. In this situation, the fluxes can also be expressed as  functions of the average fields and this results in a closed set of equations for the average fields that define a {hydrodynamic} description since it is formally equivalent to the hydrodynamic description in classical fluid mechanics \cite{CC70}. Moreover, if the spatial gradients are supposed to be small enough, the hydrodynamics is Newtonian and the resulting equations are the Navier--Stokes (NS) ones \cite{BDKS98,GS95}.
Therefore, there have been attempts to determine the NS transport coefficients of granular mixtures in the low density regime \cite{SGNT06,GD02,GVM09} and also at moderate densities \cite{JM87,JM89,Z95,AW98,WA99,GDH07,GHD07,GV09}.

On the other hand, the inelasticity in the collisions (i.e., the kinetic energy loss in interparticle collisions), introduces an inherent time scale that results, if a steady state needs to be maintained, in a minimum size of the gradients as a function of the degree of inelasticity \cite{VU09}. This renders the NS approach not appropriate for most steady rapid granular flows. In fact, we already know that, for example, the simple shear flow for a granular gas \cite{C89,C90,G03} is inherently non-Newtonian \cite{SGD04}. In addition, previous works on the latter flow  have shown that nonlinear effects can significantly modify segregation criteria with respect to NS hydrodynamic theories \cite{GV10}.

Obviously, the determination of non-Newtonian hydrodynamic profiles and transport coefficients is a prerequisite for a more complete description of transport and segregation  of impurities immersed in a granular gas. This fact motivated a previous work by the authors, in which, via a kinetic model for a multicomponent granular gas \cite{VGS07}, we determined an exact analytical solution of the problem of the steady planar Couette flow for an impurity  in a low density granular gas \cite{VGS08}.  This solution presents the advantage of including terms of all orders in the velocity and temperature spatial gradients, thus capturing nonlinear effects such as normal stress differences and a heat flux component normal to the thermal gradient.  On the other hand, recent theory results on monocomponent granular gases have revealed new and interesting classes of flows that support the validity of a hydrodynamic description, without the restriction of small spatial gradients \cite{GS96,SGN96,SG98,L04,VSG10,VGS11}. It is thus interesting (a) to check  if  the hydrodynamic profiles found in the kinetic model description are shared by the true Boltzmann equation and (b)  to gauge the degree of accuracy of the analytical solution derived in \cite{VGS08}.

\begin{figure}
\begin{center}
\includegraphics[width=.5\columnwidth]{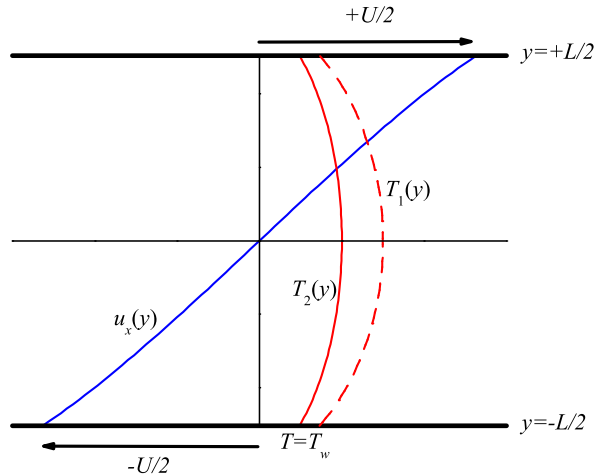}
\caption{ Sketch of the planar Couette flow. The granular gas and the impurity are
enclosed between two infinite parallel walls located at $y=\pm
L/2$, moving along the $x$-direction with velocities $\pm U/2$, and
kept at the temperature $T_w$. In the steady state, the granular gas and the impurity {move with the same flow velocity} and show different temperature profiles, $T_2(y)$ and $T_1(y)$, respectively, but with constant ratio $\chi\equiv T_1/T_2$.}
\label{fig1}
\end{center}
\end{figure}

This has motivated the present work, where by means of the direct simulation Monte Carlo  (DSMC) method \cite{B94} we obtain the numerical solution of the Boltzmann  equation associated with a smooth hard sphere impurity immersed in a low density gas of  inelastic smooth hard spheres under Couette flow. As in the case of the kinetic model, the DSMC  solution is in principle not restricted to small spatial gradients \cite{B94}.

Let us consider a set of identical smooth hard spheres (of diameter $\sigma_2$ and mass $m_2$) that collide inelastically with each other. The  inelasticity of collisions is characterized by a constant coefficient of normal restitution $\alpha_2$.  We will consider that the number density of the system is sufficiently low that the typical contact times at collisions are much smaller than the typical time interval between collisions.   In this low density regime, one can neglect the velocity correlations between the particles that are about to collide (`molecular chaos' hypothesis). In these conditions, as for an ordinary gas of hard spheres, we can describe statistically this granular gas through the Boltzmann equation conveniently adapted to take into account the inelasticity in the collisions \cite{GS95,DBS97,BP04}.  The granular gas is enclosed between two infinite parallel walls, here assumed to be both at the same temperature $T_w$, moving with a relative velocity  $U$. In this way, the sheared granular gas reaches a steady state with a non-zero velocity profile. The base laminar flows in this geometry (see figure \ref{fig1}) should be of the form $\mathbf{u}_2=u_{2,x}(y)~\mathbf{e}_x$. Moreover, a temperature profile $T_2(y)$ and a density profile $n_2(y)$ are present.  We introduce now another set of inelastic smooth hard spheres (of diameter $\sigma_1$ and mass $m_1$). The concentration of this new species  is assumed to be negligible, i.e., its density ($n_1$) relative to that of the granular gas ($n_2$) tends to zero: $x_1=n_1/n_2\to 0$ (tracer limit). For this reason we call the `impurity'  granular species number 1. The inelasticity of a collision between a sphere of species 1 and a sphere of species 2 is characterized by a constant coefficient of normal restitution $\alpha_1$, which in general differs from $\alpha_2$. Due to the shearing from the boundaries, the impurity species will also reach a base steady flow of the same type as that of the granular gas, i.e., $\mathbf{u}_1=u_{1,x}(y)~\mathbf{e}_x$, $T_1(y)$, and $n_1(y)$.

{}From the exact solution to the kinetic model mentioned above, one finds that the hydrodynamic profiles satisfy the following properties. Regarding the excess component (host granular gas) \cite{TTMGSD01}, (i) the pressure $p_2=n_2T_2$ is uniform, (ii) the (local) shear rate $\partial u_{2,x}/\partial y$ scaled with respect to the (local) collision frequency $\nu_2\propto n_2 T_2^{1/2}$ is uniform, and (iii) the temperature $T_2$ is a parabolic function of the flow velocity $u_{2,x}$. In the case of the tracer component (impurity), one finds that \cite{VGS08} (iv) the flow velocity coincides with that of the gas, i.e., $u_{1,x}=u_{2,x}$, (v) the mole fraction $n_1/n_2$ is uniform, and (vi) the temperature ratio $\chi\equiv T_1/T_2$ is also uniform. As we said, the main goal of the paper is to confirm these predictions by means of computer simulations. In addition, we will measure the momentum and heat fluxes of the impurity to get the generalized transport coefficients and compare them with the analytical results derived from the kinetic model.

The structure of the paper is the following. We formally describe the kinetic theory problem in section \ref{sec2}, where the generalized rheological and transport coefficients are also defined. In this section \ref{sec2} we also briefly recall the  kinetic model results and describe the numerical method (DSMC). In section \ref{sec3} we present the simulation data compared with the analytical solution of the kinetic model and discuss the results. Finally, we present the conclusions in section \ref{sec4}. Additionally, we present in the appendix the theoretical hydrodynamic properties for the impurity from our previous work \cite{VGS08}.

\section{Theoretical description and numerical methods\label{sec2}}

\subsection{The Boltzmann description of the Couette flow for the granular gas and the impurity \label{Boltzmann}}
We consider  a granular gas composed of inelastic $d$-dimensional hard spheres of diameter $\sigma_2$, mass $m_2$, and coefficient of normal restitution $\alpha_2$. In the low density regime, the corresponding velocity distribution function $f_2(\mathbf{r},\mathbf{v},t)$  (in the absence of gravity) obeys the Boltzmann equation
\begin{equation}
\left(\partial_t +\mathbf{v}\cdot\nabla\right) f_2=J_{22}[\mathbf{v}|f_2,f_2],
\label{Be2}
\end{equation}
{where  $J_{22}[\mathbf{v}|f_2,f_2]$ is the (inelastic) collision operator.}

We add now the impurity particles {of diameter $\sigma_1$ and mass $m_1$} (species 1), which are present in a vanishing concentration.
For this reason, we can assume that collisions between impurity particles themselves may be neglected and,
in addition, the state of the granular gas (species 2) is not affected by the presence of impurities, so  equation (\ref{Be2}) is still valid. The velocity distribution function
$f_1(\mathbf{r},\mathbf{v},t)$ for the impurity particles obeys the Boltzmann--Lorentz equation
\begin{equation}
\left(\partial_t +\mathbf{v}\cdot\nabla\right) f_1=J_{12}[\mathbf{v}|f_1,f_2], \label{Be1}
\end{equation}
{where  $J_{12}[\mathbf{v}|f_1,f_2]$ is the corresponding (inelastic) collision operator, which is parameterized by the impurity-gas coefficient of normal restitution $\alpha_1$. Detailed expressions for $J_{ij}[\mathbf{v}|f_i,f_j]$ can be found in, for instance,  \cite{VGS07}. }

The relevant hydrodynamic fields for both species are the number densities $n_i$, flow velocities $\mathbf{u}_i$, and  granular temperatures $T_i$. They are defined by the relations
\beq
n_i(\mathbf{r},t)=\int\mathrm{d} \mathbf{v}\, f_i(\mathbf{r},\mathbf{v},t),
\eeq
\beq
\mathbf{u}_i(\mathbf{r},t)=\frac{1}{n_i(\mathbf{r},t)}\int\mathrm{d} \mathbf{v} \,\mathbf{v}f_i(\mathbf{r},\mathbf{v},t),
\eeq
\beq
T_i(\mathbf{r},t)=\frac{m_i}{dn_i(\mathbf{r},t)}\int\mathrm{d} \mathbf{v}\, \left[\mathbf{v}-\mathbf{u}_i(\mathbf{r},t)\right]^2 f_i(\mathbf{r},\mathbf{v},t).
\label{fields}
\eeq
In equation \eref{fields} we have defined the partial temperatures $T_i$ taking the velocities of species $i$ relative to its mean value $\mathbf{u}_i$. The usual choice, however, is to refer the velocities to the global mean flow velocity $\mathbf{u}$ \cite{VGS08}. Here, for convenience, we adopt the former choice. In any case, since $\mathbf{u}=\mathbf{u}_2$ in the tracer limit, both choices are equivalent in the case of the granular gas.
Additionally, the pressure tensor $\mathsf{P}_i$ and the heat flux $\mathbf{q}_i$ for each species can be defined as
\beq
\mathsf{P}_i(\mathbf{r},t)=m_i\int\mathrm{d} \mathbf{v}\, \left[\mathbf{v}-\mathbf{u}_i(\mathbf{r},t)\right]\left[\mathbf{v}-\mathbf{u}_i(\mathbf{r},t)\right]f_i(\mathbf{r},\mathbf{v},t),
\eeq
\beq
\mathbf{q}_i(\mathbf{r},t)=\frac{m_i}{2}\int\mathrm{d} \mathbf{v}\, \left[\mathbf{v}-\mathbf{u}_i(\mathbf{r},t)\right]^2\left[\mathbf{v}-\mathbf{u}_i(\mathbf{r},t)\right]f_i(\mathbf{r},\mathbf{v},t).
\eeq

Given that the microscopic state of the granular gas is independent of the microscopic state of the impurities, the mass, momentum, and energy balance equations for species 2 take the usual form \cite{BDKS98}
\beq
D_t n_2+n_2\nabla\cdot \mathbf{u}_2=0,
\label{nbal}
\eeq
\beq
D_t\mathbf{u}_2+\frac{1}{m_2n_2}\nabla\cdot\mathsf{P}_2=\mathbf{0},
\label{ubal}
\eeq
\beq
D_tT_2+\frac{2}{dn_2}\left(\nabla\cdot\mathbf{q}_2+\mathsf{P}_2:\nabla
\mathbf{u}_2\right)=-\zeta_2 T_2,
\label{Tbal}
\eeq
where  $D_t\equiv\partial_t+\mathbf{u}_2\cdot\nabla$ is the material
time derivative and
\beq
\zeta_2=-\frac{m_2}{dn_2}\int\dd \mathbf{v}\, v^2 J_{22}[\mathbf{v}|f_2,f_2]
\label{zeta2}
\eeq
is the cooling rate of the granular gas.

Now we assume that the system (granular gas plus impurities) is subject to the planar Couette flow. For steady states ($\partial_t=0$), and given the  geometry of the problem ($\partial_x=\partial_z=0$), we obtain
\beq
\partial_yP_{2,xy}=\partial_yP_{2,yy}=0,
 \label{Pbaly}
\eeq
\beq
\partial_yq_{2,y}+P_{2,xy}\partial_y u_{2,x}=-\frac{d}{2}\zeta_2 n_2T_2 .
\label{Tbaly}
\eeq

Thus far, all the equations in this section are formally exact in the framework of the Boltzmann equation. Based on previous results \cite{VGS08,TTMGSD01} derived from the kinetic model described below, we {\emph{expect}} that the hydrodynamic fields of the gas in the bulk domain of the system  have the forms
\begin{equation}
\label{III.5}
p_2=n_2T_2=\text{const} ,
\end{equation}
\begin{equation}
\label{III.6}
\frac{1}{\nu_2}\partial_yu_{2,x}=a=\text{const},
\end{equation}
\begin{equation}
\label{III.7}
\frac{1}{2m_2\text{Pr}}\left(\frac{1}{\nu_2}\partial_y\right)^2T_{2}=- \gamma=\text{const}.
\end{equation}
Here, $\mathrm{Pr}=(d-1)/d$ is the conventional Prandtl number \cite{CC70} and $\nu_2\propto n_2 T_2^{1/2}$ is a characteristic collision frequency. For the sake of concreteness, henceforth we will take
\begin{equation}
\nu_2\equiv \frac{8\pi^{(d-1)/2}}{(d+2)\Gamma(d/2)}n_2\sigma_2^{d-1}\left(T_2/m_2\right)^{1/2}.
\label{nu2}
\end{equation}

In equation \eref{III.6}, the constant $a$ represents a dimensionless shear rate. It plays the role of a Knudsen number associated with the velocity gradient. The constant  $\gamma$ in equation \eref{III.7} is a dimensionless parameter  {(henceforth called thermal curvature
coefficient)} characterizing the curvature of the temperature
{profile} as a consequence of both the viscous heating and the
collisional cooling. As a consequence, $\gamma$ must depend on both the shear rate $a$ and the coefficient of restitution
$\alpha_{2}$. It is interesting to note that, from equations \eref{III.6} and \eref{III.7}, one finds
\beq
T_2(y)=T_2(0)-\text{Pr}\frac{m_2\gamma}{a^2}u_{2,x}^2(y),
\label{EOS}
\eeq
where we have considered that $u_{2,x}(0)=0$ and the temperature profile is symmetric.
Equation \eref{EOS} implies that, if eliminating $y$ between $T_2$ and $u_{2,x}$, the temperature is a linear function of $u_{2,x}^2$.

With respect to the state of the impurities, we assume that it is enslaved to that of the gas. This means that \cite{VGS08}
\beq
\mathbf{u}_1 =\mathbf{u}_2,
\label{u1}
\eeq
\beq
x_1\equiv \frac{n_1}{n_2}=\text{const},
\label{n1}
\eeq
\beq
\chi\equiv \frac{T_1}{T_2}=\text{const}.
\label{chi}
\eeq
The hypothesis \eref{u1} implies that there is no diffusion of the impurities with respect to the gas particles. Equations \eref{n1} and \eref{chi}, together with equation \eref{III.5}, imply that $p_1=n_1T_1=\text{const}$. In summary,  the hydrodynamic profiles of the system are provided by equations \eref{III.5}--\eref{III.7} and \eref{u1}--\eref{chi}.

In order to characterize the non-Newtonian properties, it is convenient to introduce the following five dimensionless rheological factors \cite{VGS08,VSG10,VGS11,TTMGSD01}
\beq
{P_{i,xy}=-\eta_i^*(a)\frac{n_iT_i}{\nu_i}\frac{\partial u_{i,x}}{\partial y}},
\label{Pxy}
\eeq
\beq
\theta_{i,x}(a)=\frac{P_{i,xx}}{n_i T_i},\qquad \theta_{i,y}(a)=\frac{P_{i,yy}}{n_i T_i},
\label{thetas}
\eeq
\beq
{q_{i,y}=-\lambda_i^*(a)\frac{d+2}{2m_i\mathrm{Pr}}\frac{n_iT_i}{\nu_i}\frac{\partial T_i }{\partial y},}
\label{qy}
\eeq
\beq
{q_{i,x}=\phi_i^*(a)\frac{d+2}{2m_i\mathrm{Pr}}\frac{n_iT_i}{\nu_i}\frac{\partial T_i }{\partial y},}
\label{qx}
\eeq
where  $\nu_2$ is defined by equation \eref{nu2}  and
\beq
\nu_1\equiv
\frac{4\pi^{(d-1)/2}}{(d+2)\Gamma(d/2)}n_2\sigma_{12}^{d-1}\left(\frac{2T_1}{m_1}+\frac{2T_2}{m_2}\right)^{1/2}.
\label{nu1}
\eeq
{Note that, in the case $i=1$, the definitions of $\eta_i^*$, $\lambda_i^*$, and $\phi_i^*$  in equations \eref{Pxy}, \eref{qy}, and \eref{qx} slightly differ from those in \cite{VGS08}}. The nondimensionalization of the generalized shear viscosity $\eta_i^*$ and thermal conductivity $\lambda_i^*$ is such that $\eta_2^*=\lambda_2^*=1$ for an elastic gas ($\alpha_2=1$) in the NS regime ($a\to 0$) \cite{CC70}. Note that, while $\eta_i^*$ and $\lambda_i^*$ are generalizations of NS transport coefficients, the functions $\theta_{i,x}$, $\theta_{i,y}$, and $\phi_i^*$ are generalizations of Burnett transport coefficients \cite{CC70,B34,LR03}.

Of course, if the impurities are mechanically equivalent to the gas particles (i.e., $m_1=m_2$, $\sigma_1=\sigma_2$, and $\alpha_1=\alpha_2$), the transport coefficients associated with both species coincide.

Taking into account the constitutive forms \eref{Pxy} and \eref{qy} for the granular gas ($i=2$), as well as equations \eref{III.5}--\eref{III.7}, the exact balance equation \eref{Tbaly} becomes
\beq
\eta_2^*a^2-(d+2)\lambda_2^*\gamma=\frac{d}{2}\zeta_2^*, \qquad \zeta_2^*\equiv \frac{\zeta_2}{\nu_2}.
\label{eta2}
\eeq
Note that in the elastic case ($\zeta_2^*=0$) and in the NS limit (i.e., $\eta_2^*\to 1$ and $\lambda_2^*\to 1$) one has $\gamma=a^2/(d+2)$ \cite{GS03}.
In general, $\gamma$ depends on both $a$ and $\alpha_2$, its sign depending on the competition between viscous heating and inelastic cooling. If viscous heating dominates (i.e., $\eta_2^*a^2>d\zeta_2^*/2$), then $\gamma>0$. On the other hand, $\gamma<0$ in the opposite situation (i.e., $\eta_2^*a^2<d\zeta_2^*/2$). Both effects cancel each other (and thus $\gamma=0$) at a threshold shear rate $a_\kk$ given by
\beq
a_\kk^2=\frac{d}{2}\frac{\zeta_2^*}{\eta_2^*(a_\kk)}.
\label{ath2}
\eeq
Since $\gamma=0$ at $a=a_\kk$, it follows from equations \eref{III.7} and \eref{chi} that $\nu_i^{-1}\partial_y T_i=\text{const}$. This condition, together with $n_iT_i=\text{const}$ and equations \eref{qy} and \eref{qx}, implies that the heat flux is uniform at   $a=a_\kk$ \cite{VSG10,VGS11,SGV09} both for the impurity and the host gas.

\subsection{The kinetic model description of the Couette flow for the granular gas and the impurity} \label{BGK}

The mathematical complexity of the Boltzmann {and Boltzmann--Lorentz} equations \eref{Be2} and {\eref{Be1}} prevents one from obtaining exact solutions. This has motivated the proposal of simpler kinetic models, most of them inspired on the well-known Bhatnagar--Gross--Krook (BGK) kinetic model for ordinary gases \cite{C88}. Here we consider the following BGK-type kinetic model {\cite{VGS07,BDS99}}:
\beq
{J_{i2}[\mathbf{v}|f_i,f_2]\to - k_d\frac{1+\alpha_i}{2}\nu_{i}
\left(f_i-f_{i2}\right)+\frac{\zeta_{i}}{2}\frac{\partial
}{\partial {\bf v}}\cdot \left[\left( {\bf v
}-\mathbf{u}_{i}\right)f_i\right]},
\label{BGK2}
\end{equation}
where
\begin{equation}
{ f_{i2}(\mathbf{v})=n_i\left(\frac{m_i}{2\pi
T_{i2}}\right)^{d/2}
\exp\left[-\frac{m_i}{2T_{i2}}\left(\mathbf{v}-\mathbf{u}_{i2}\right)^2\right]}
\label{4.12b}
\end{equation}
is a reference distribution function. {In the case of the granular gas ($i=2$), $T_{22}=T_2$ and $\mathbf{u}_{22}=\mathbf{u}_{2}$, so $f_{22}$ is the local equilibrium distribution function. In the case of the impurity particles ($i=1$) \cite{VGS07},}
\beq
\fl
T_{12}=T_1+\frac{2\mu}{(1+\mu)^2}\left\{T_2-T_1+\frac{(\mathbf{u}_1-\mathbf{u}_2)^2}{2d}\left[m_2+
\frac{T_2-T_1}{T_1/m_1+T_2/m_2}\right]\right\},
\label{4.18}
\end{equation}
\begin{equation}
\mathbf{u}_{12}=\frac{\mu\mathbf{u}_{1}+\mathbf{u}_{2}}{1+\mu},
\label{4.16}
\end{equation}
{where}
\beq
\mu\equiv \frac{m_1}{m_2}
\label{mu}
\eeq
is the mass ratio.
{In equation \eref{BGK2}}
\beq
\zeta_2=\frac{d+2}{4d}(1-\alpha_2^2)\nu_2
\label{0.1a}
\eeq
is the cooling rate \eref{zeta2} evaluated in the local equilibrium approximation, {while}
\begin{equation}
\label{zetasij}
\zeta_{1}=\frac{d+2}{2d}\frac{\nu_{1}}{(1+\mu)^2}\left[1+\frac{m_1{T}_2}{m_2
{T}_1}+\frac{3}{2d}\frac{m_1}{{T}_1}\left({\bf
u}_1-{\bf u}_2\right)^2\right](1-\alpha_{1}^2)
\end{equation}
is the impurity cooling rate. Finally, the factor $k_d$ can be chosen to optimize agreement with the Boltzmann {description}. In particular, the choices $k_d=1$ and $k_d=\text{Pr}=(d-1)/d$ reproduce  the NS shear viscosity and  thermal conductivity coefficients, respectively, of the {gas in the elastic limit} \cite{CC70,C88}.
A third criterion to fix the factor $k_d$ is to require that the collisional momentum transfer of the impurities be the same for the kinetic model as for the true Boltzmann--Lorentz equation \cite{VGS07,VGS08}. This results in $k_d=(d+2)/d$.

In a recent work \cite{VGS08}, we solved the kinetic model equations for the granular gas and impurity in the steady Couette flow. The resulting profiles agree with the forms \eref{III.5}--\eref{III.7} and \eref{u1}--\eref{chi}. Moreover, the solution gives $\gamma$, $\chi$, $\eta_1^*$, $\theta_{1,x}$, $\theta_{1,y}$, $\lambda_1^*$, and $\phi_1^*$ as functions of $a$, $\alpha_1$, $\alpha_2$, $\mu$, and $\omega\equiv \sigma_1/\sigma_2$. Their  explicit forms are displayed in the appendix.

\subsection{Numerical methods (Monte Carlo simulations)} \label{sec23}

\begin{figure}\begin{center}
\includegraphics[width=.5\columnwidth]{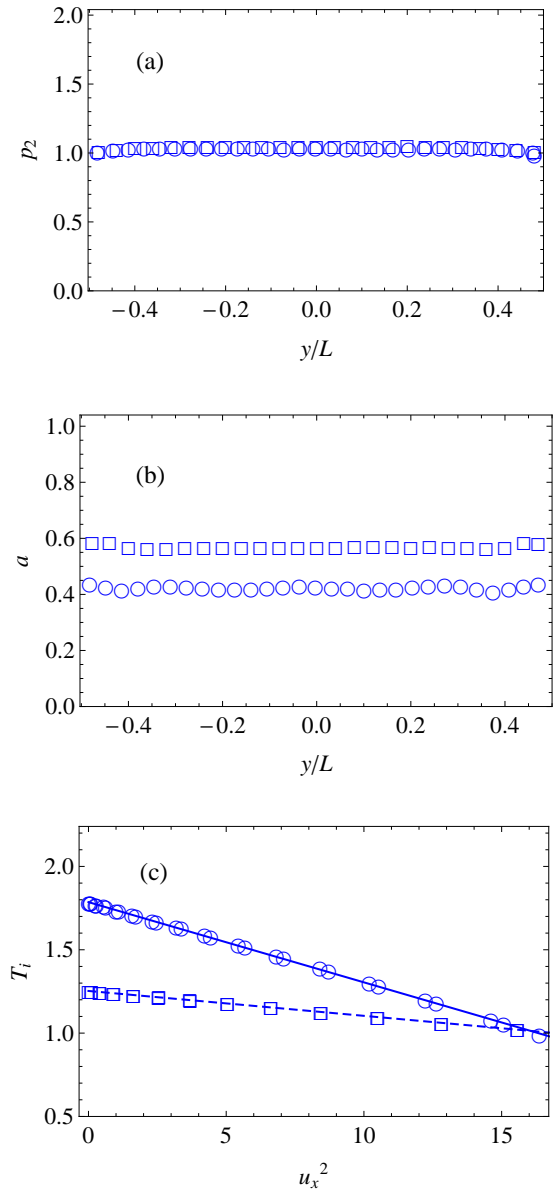}
\caption{{This figure illustrates the verification in the bulk region of the three hypotheses on the stationary Couette flow for the granular gas: (a) the hydrostatic pressure $p_2$ is constant, (b) the local shear rate $a$ is also constant, and (c) the temperature $T_2$ is a linear function of $u_{2,x}^2$. Two values of the coefficient of restitution are considered:  $\alpha_2=0.9$  ($\bigcirc$, with $L=23.23$ and $a=0.443$) and $\alpha_2=0.8$ ($\square $, with $L=15.48$ and $a=0.571$). Lines in (c) stand for linear fits to DSMC data.}}
\label{gr3p1}
\end{center}\end{figure}

\begin{figure}\begin{center}
\includegraphics[width=.5\columnwidth]{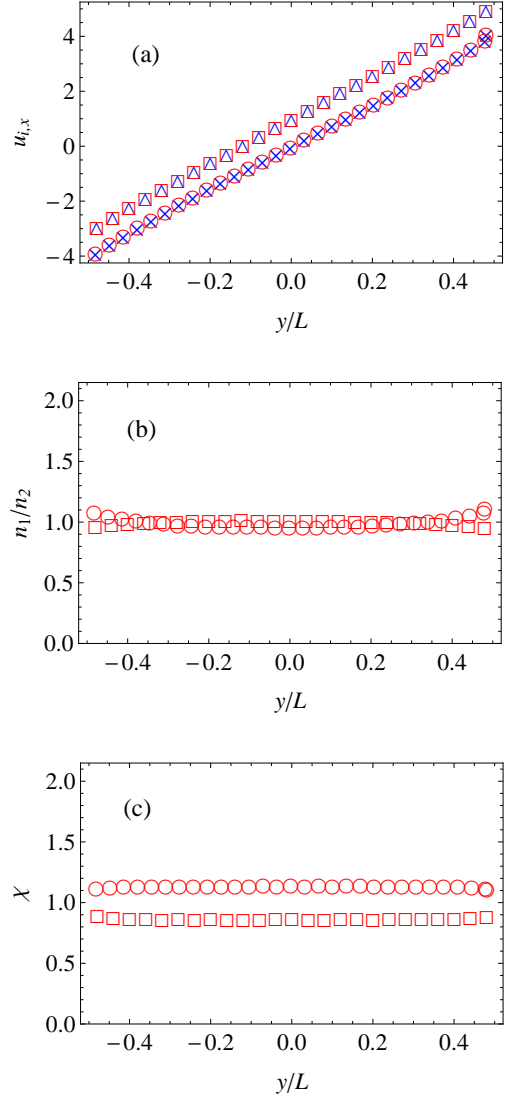}
\caption{{This figure illustrates the verification in the bulk region of the three hypotheses on the stationary Couette flow for the impurity: (a) no mutual diffusion exists, i.e., $u_{1,x}(y)=u_{2,x}(y)$, (b) the mole fraction $n_1/n_2$  is constant, and (c) the temperature ratio $\chi=T_1/T_2$ is also constant. Two cases are considered:  $\alpha_1=\alpha_2=0.9$, $\mu=2$, $\omega=1$ ($\bigcirc$, with $L=23.23$ and $a=0.443$) and  $\alpha_1=\alpha_2=0.8$, $\mu=0.5$, $\omega=1$ ($\square $, with $L=15.48$ and $a=0.571$). In panel (a)  the crosses and triangles correspond to the granular gas ($i=2$), and the symbols corresponding to $\alpha=0.8$ have been lifted to avoid overlap with the symbols corresponding to $\alpha=0.9$. In panel (b) the density $n_i$ is normalized with respect to its spatial average value $\overline{n}_i$.}}
\label{gr3p2}
\end{center}\end{figure}

As said in section \ref{sec1}, we have solved numerically the Boltzmann and Boltzmann--Lorentz kinetic equations for the hard-sphere ($d=3$) granular gas and impurity [equations \eref{Be2} and \eref{Be1}, respectively] by means of the DSMC method. Originally devised for elastic gases in the low density limit \cite{B94}, this method has been successfully  implemented also for inelastic hard spheres \cite{VSG10} and moderately dense gases (Enskog kinetic equation \cite{F97,MS97,MGSB99}).

The implementation of the algorithm for granular gases has been described in more detail elsewhere (see, for instance,  \cite{VGS11}).
We will recall only that it consists of small time steps (in the scale of the characteristic mean free time), each one having two basic stages: (i) free streaming and (ii) {stochastic} interparticle (binary) collisions. The system is divided into small cells (in the scale of the characteristic mean free path). {In the DSMC method, the number $N_i$ of simulated particles is a statistical technical parameter that, in contrast to the molecular dynamics case, does not need to coincide with the actual number of physical particles in the system.  We have used the same number of simulated particles ($N_1=N_2=2\times 10^5$) for both species. On the other hand, since the physical impurity concentration is assumed to be negligibly small, only 2-2 collisions are considered in the evolution of the granular gas and only 1-2 collisions are considered in the evolution of the impurity particles. Therefore, after a collision of type 2-2, the velocities of both particles are changed, while only the velocity of the impurity particle is changed after a collision of type 1-2. }

As in a previous work \cite{VGS11}, we  perform two averages for steady states: (i) a first spatial average over neighbor simulation cells, taking care that this coarse-grained cell size is not larger than the typical length over which hydrodynamic fields vary, and (ii) a time average for each coarse-grained cell since the microstates of the simulation are stored iteratively many times during the stationary state  in the simulation.

Since the  parameter space (mass ratio $\mu\equiv m_1/m_2$, size ratio $\omega\equiv \sigma_1/\sigma_2$, coefficients of restitution $\alpha_1$ and $\alpha_2$, and shear rate $a$) is quite large, we have focused on a few representative cases. As in our previous work on the BGK-type model \cite{VGS08}, we have analyzed cases with a common coefficient of restitution ($\alpha_1=\alpha_2=\alpha$) and equal diameter ($\omega=1$) for $d=3$ (spheres). Three different values of the mass ratio ($\mu=2, 1, 0.5$) and of the coefficient of restitution ($\alpha=1, 0.9, 0.8$) have been considered. Thus, the elastic limit  ($\alpha=1$) as well as the properties of the granular gas ($\mu=1$) are included as particular cases. For each one of  the nine combinations of the pair $(\mu,\alpha)$, we have made series of simulations in shear rate $a$ by keeping fixed the wall velocity difference $U=10$, but varying the wall distance in the range $L=2.5$--$30$. We will use hereafter nondimensionalized quantities with the following choice of units: $m_2=1$, $T_2(\pm L/2)=1$, and $\nu_2(\pm L/2)=1$. Also, we define the density levels $\overline{n}_i=1$, where the bar denotes a spatial average across the system.

In section \ref{sec3} we compare the DSMC numerical solution of the Boltzmann description with the analytical solution of the BGK-type kinetic model. But, before that, it is appropriate  to show DSMC data confirming that the hypotheses on which our theoretical description relies on [equations \eref{III.5}, \eref{III.6}, \eref{EOS}--\eref{chi}] are indeed valid.
As an illustration,  figures \ref{gr3p1} and \ref{gr3p2} display the DSMC profiles for the cases $(\mu,\alpha)=(2,0.9)$ and $(\mu,\alpha)=(0.5,0.8)$. {}From figure \ref{gr3p1} we observe that the hydrostatic pressure $p_2$ of the granular gas, except for small inflections near the boundary layers, is flat, the local shear rate $a$ is indeed constant throughout the system, and the temperature $T_2$ is a linear function of $u_{2,x}^2$. Figure \ref{gr3p2}  confirms the remaining hypotheses \eref{u1}--\eref{chi}, specific for the granular impurity. The ``enslaving'' condition $\mathbf{u}_1=\mathbf{u}_2$ is fulfilled with a very high degree of accuracy for all points in the system, even in the boundary layers. Moreover, the relative impurity concentration and the temperature ratio are practically constant.

\section{Results and discussion}\label{sec3}

\subsection{The threshold shear rate}

\begin{figure}\begin{center}
\includegraphics[width=.5\columnwidth]{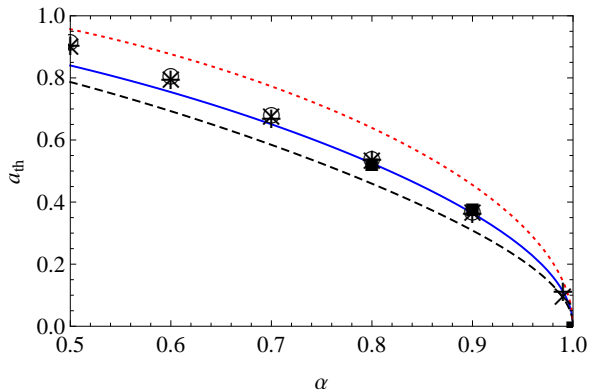}
\caption{{Threshold values $a_{\kk}$ for the shear rate $a$ vs the coefficient of normal restitution for spheres ($d=3$). The lines represent the theoretical  results given by equation \protect\eref{III.15} with three different choices for the parameter $k_d$:  $k_d=1$ (solid line),  $k_d=\mathrm{Pr}=(d-1)/d$ (dashed line), and $k_d=(d+2)/d$ (dotted line). Symbols stand for DSMC simulation data with  $\Delta T=0$ ($\blacksquare$, this work),  $\Delta T=2$ ($\times$, \protect\cite{VSG10,VGS11}), $\Delta T=10$ ($+$, \protect\cite{VSG10,VGS11}), and simple shear flow ($\bigcirc$, \protect\cite{AS05}).}}
\label{ath}
\end{center}\end{figure}

{We first consider} the threshold value of the shear rate $a_\text{th}$ at which the thermal curvature parameter $\gamma$ vanishes. As discussed below equation \eref{eta2}, the value $a=a_\kk$ is especially important since it corresponds to an exact balance between viscous heating and inelastic cooling, giving rise to $\mathbf{q}_i=\text{const}$. In the geometry sketched in figure \ref{fig1}, where both walls are maintained at the same temperature, the value $\gamma=0$ implies a constant temperature $T_2$ and thus the Couette flow becomes equivalent to the well-known simple shear flow \cite{C89,C90,G03}. More in general, when the walls are allowed to have different temperatures ({$\Delta T=T_{w+}/T_{w-}-1\neq 0$}), the limit case $\gamma=0$ defines, in the parameter space \{$\alpha, a, \Delta T$\}, a surface that has been shown recently \cite{VSG10,VGS11} to represent a special class  of granular flows, including the conventional Fourier flow of an elastic gas. This generalized class of flows (that we called ``LTu'' because the temperature is a linear function of the flow velocity) can be theoretically described in a single hydrodynamic theory frame, both for elastic and granular gases.

Figure \ref{ath} shows the $\alpha$-dependence of the threshold shear rate $a_\text{th}$, as predicted by the BGK-type kinetic model, equation \eref{III.15}, with the three choices of $k_d$ mentioned before, namely {$k_d=1$, $k_d=(d-1)/d$, and $k_d=(d+2)/d$}. The simulation data obtained here for $\alpha=0.8$ and $0.9$ with $\Delta T=0$, as well as those of  \cite{VGS11} for a wider range with $\Delta T\neq 0$, are also included in figure \ref{ath}. We clearly observe that the best agreement is achieved with the choice $k_d=1$. {Although we have checked that either $k_d=(d-1)/d$ or $k_d=(d+2)/d$ may provide a better agreement with simulation for some of the other quantities,   henceforth we will adopt the choice $k_d=1$ as a convenient compromise between simplicity and accuracy.}

In the rest of this section, we will compare DSMC results from the Boltzmann equation with the analytical results derived from our BGK-type model, representing the relevant hydrodynamic properties as  functions of the shear rate {(for $a\geq a_\kk$)}, so we can analyze to what extent the nonlinear theoretical description is reliable, {at least at a qualitative level,}  for these very far away from equilibrium steady states.

\subsection{Thermal curvature coefficient $\gamma$ and temperature ratio $\chi$}
\begin{figure}\begin{center}
\includegraphics[width=.5\columnwidth]{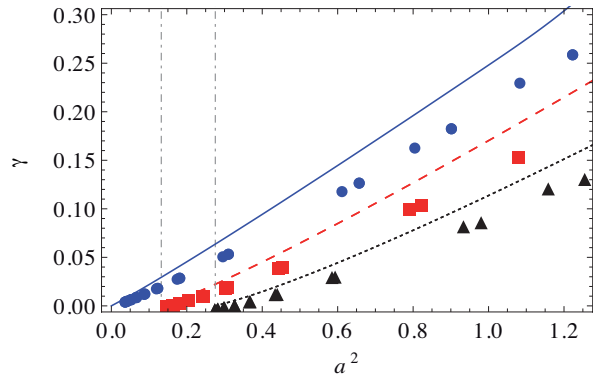}
\caption{{Thermal curvature coefficient $\gamma$ as a function of the square of the shear rate, $a^2$. From now on, in all figures, lines stand for the BGK-type theory (with $k_d=1$) and symbols stand for DSMC data. Three series are plotted: $\alpha=1$ (solid line and circles), $\alpha=0.9$ (dashed line and squares), and $\alpha=0.8$ (dotted line and triangles). The threshold levels $a_{\kk}^2$ for the shear rate in the cases $\alpha=0.9$ and $\alpha=0.8$ are marked with dotted-dashed vertical lines.}}
\label{gamma}
\end{center}\end{figure}

\begin{figure}\begin{center}
\includegraphics[width=.5\columnwidth]{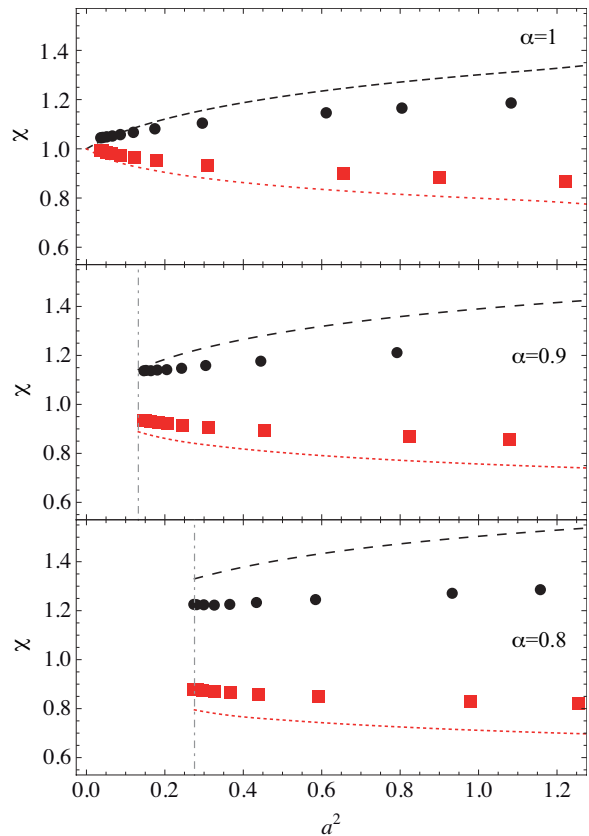}
\caption{{Temperature ratio $\chi\equiv T_1/T_2$ as a function of the square of the shear rate, $a^2$.  Two mass ratios are considered for each value of $\alpha$:  $\mu\equiv m_1/m_2=2$ (dashed lines and circles) and $\mu=1/2$ (dotted lines and squares). The threshold levels $a_{\kk}^2$ for the shear rate in the cases $\alpha=0.9$ and $\alpha=0.8$ are marked with dotted-dashed vertical lines.}}
\label{ratio}
\end{center}\end{figure}

In figure \ref{gamma} we can see that the theoretical thermal curvature parameter   shows a good agreement with DSMC data. The agreement is quantitatively very good  near the threshold $a_{\kk}$, but the theory tends to overestimate $\gamma$ as the shear rate increases.  {In any case, we} can  observe in figure \ref{gamma}  the reliability of our non-Newtonian hydrodynamic description in the context of the BGK-type kinetic model. As usual, the thermal curvature parameter decreases for decreasing shear rate, until it reaches a threshold value $\gamma=0$  at which we obtain the simple shear flow \cite{G03} or, more generally,  the  LTu class flow \cite{VSG10,VGS11,SGV09}. We should recall that the BGK solution is not mathematically well defined for states with $\gamma<0$ {(see, however,  \cite{VGS08b} for an analytical continuation).} In principle, it should be also possible to find nonlinear Couette flows from the Boltzmann equation (it has already been shown that these states do exist in the quasielastic limit \cite{VU09}). This region would correspond to  points below the LTu surface \cite{VSG10,VGS11}.

In figure \ref{ratio} we plot the results for the temperature ratio. We may confirm that the Boltzmann equation solution follows the same general trends as the BGK-type model \cite{VGS08}. In particular, the impurity has a higher (lower) temperature than the granular gas if its mass is larger (smaller) than that of a gas particle, this effect being more pronounced as the shear rate increases. {On the other hand, the BGK-type model  tends to exaggerate these effects.}

\subsection{Generalized transport coefficients}

\begin{figure}\begin{center}
\includegraphics[width=.5\columnwidth]{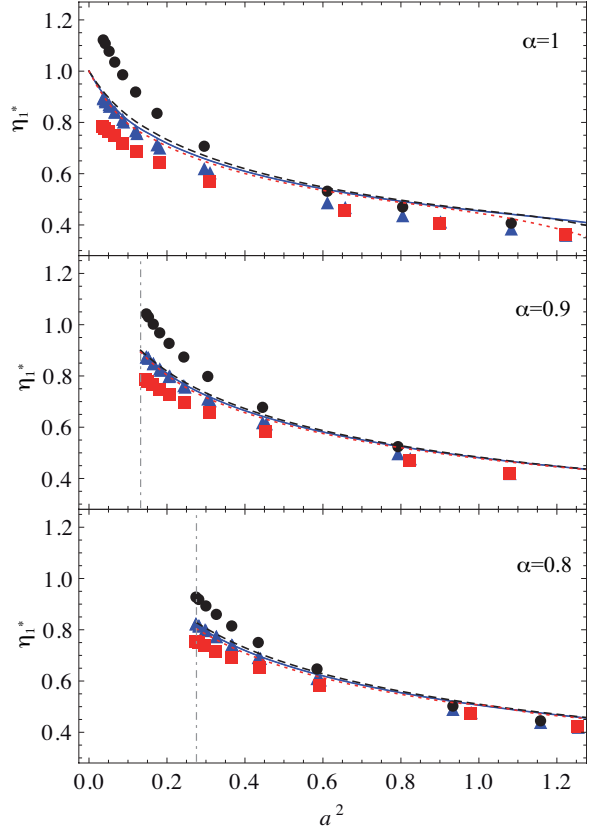}
\caption{{Reduced shear viscosity $\eta_1^*$ as a function of the square of the shear rate, $a^2$. Three mass ratios are considered for each value of $\alpha$:  $\mu\equiv m_1/m_2=2$ (dashed lines and circles), $\mu=1$ (solid lines and triangles), and $\mu=1/2$ (dotted lines and squares). The threshold levels $a_{\kk}^2$ for the shear rate in the cases $\alpha=0.9$ and $\alpha=0.8$ are marked with dotted-dashed vertical lines.}}
\label{eta}
\end{center}\end{figure}

\begin{figure}\begin{center}
\includegraphics[width=.5\columnwidth]{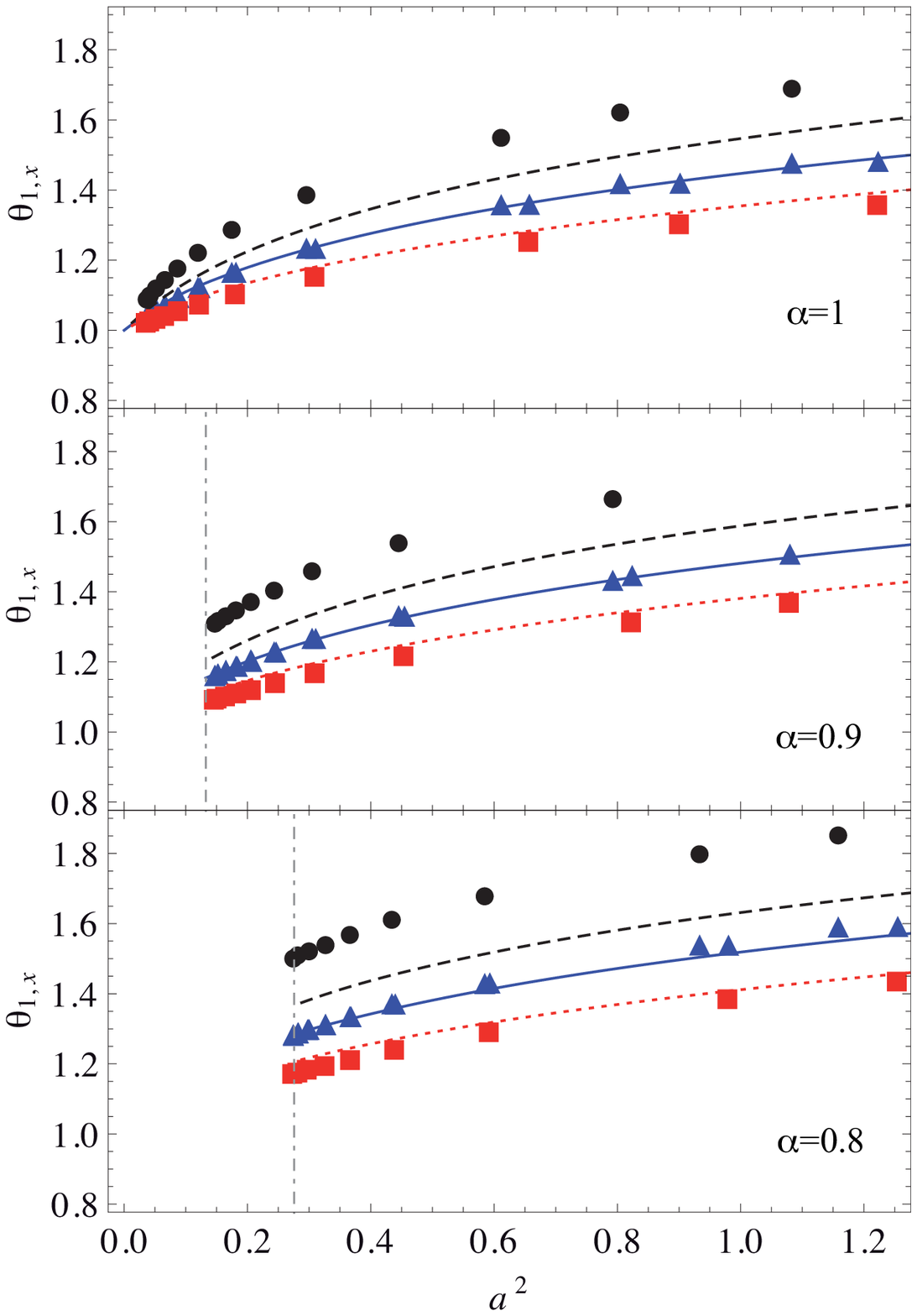}
\caption{{Normal stress coefficient $\theta_{1,x}$ as a function of the square of the shear rate, $a^2$. Three mass ratios are considered for each value of $\alpha$:  $\mu\equiv m_1/m_2=2$ (dashed lines and circles), $\mu=1$ (solid lines and triangles), and $\mu=1/2$ (dotted lines and squares). The threshold levels $a_{\kk}^2$ for the shear rate in the cases $\alpha=0.9$ and $\alpha=0.8$ are marked with dotted-dashed vertical lines.}}
\label{thetax}
\end{center}\end{figure}

\begin{figure}\begin{center}
\includegraphics[width=.5\columnwidth]{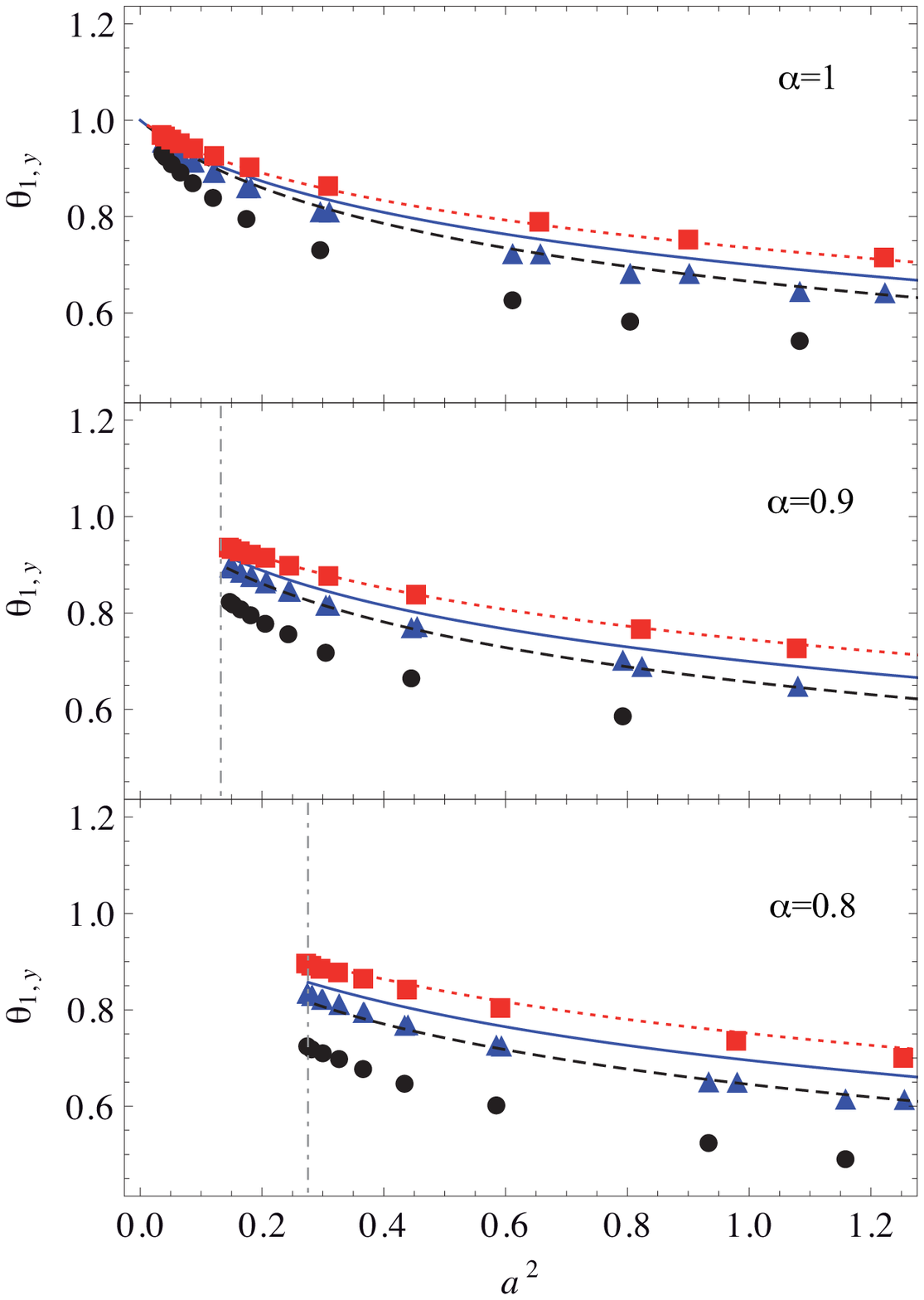}
\caption{{Normal stress coefficient $\theta_{1,y}$ as a function of the square of the shear rate, $a^2$. Three mass ratios are considered for each value of $\alpha$:  $\mu\equiv m_1/m_2=2$ (dashed lines and circles), $\mu=1$ (solid lines and triangles), and $\mu=1/2$ (dotted lines and squares). The threshold levels $a_{\kk}^2$ for the shear rate in the cases $\alpha=0.9$ and $\alpha=0.8$ are marked with dotted-dashed vertical lines.}}
\label{thetay}
\end{center}\end{figure}

\begin{figure}\begin{center}
\includegraphics[width=.5\columnwidth]{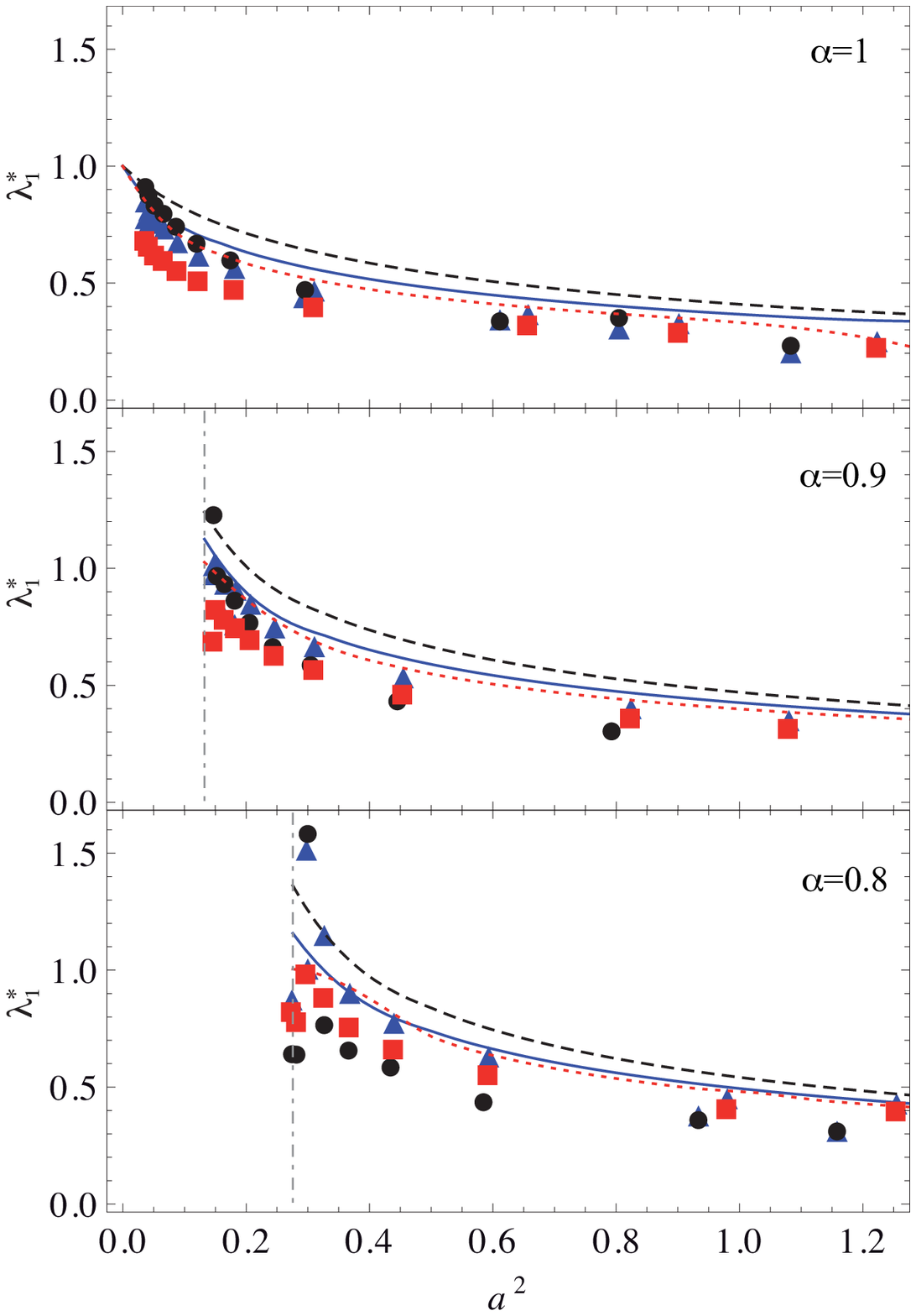}
\caption{{Reduced thermal conductivity $\lambda_1^*$ as a function of the square of the shear rate, $a^2$. Three mass ratios are considered for each value of $\alpha$:  $\mu\equiv m_1/m_2=2$ (dashed lines and circles), $\mu=1$ (solid lines and triangles), and $\mu=1/2$ (dotted lines and squares). The threshold levels $a_{\kk}^2$ for the shear rate in the cases $\alpha=0.9$ and $\alpha=0.8$ are marked with dotted-dashed vertical lines.}}
\label{lambda}
\end{center}\end{figure}

\begin{figure}\begin{center}
\includegraphics[width=.5\columnwidth]{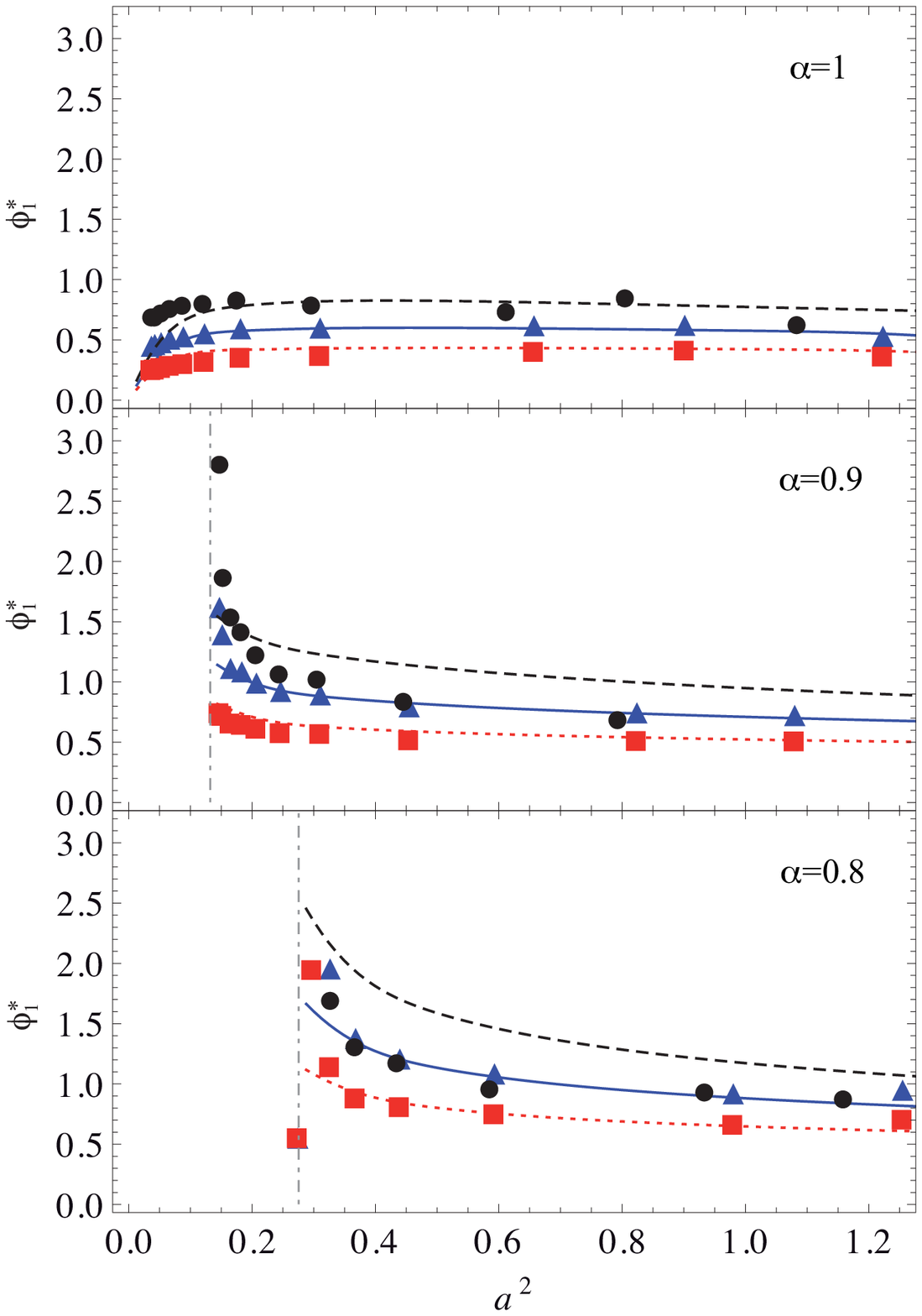}
\caption{{Reduced cross thermal conductivity $\phi_1^*$ as a function of the square of the shear rate, $a^2$. Three mass ratios are considered for each value of $\alpha$:  $\mu\equiv m_1/m_2=2$ (dashed lines and circles), $\mu=1$ (solid lines and triangles), and $\mu=1/2$ (dotted lines and squares). The threshold levels $a_{\kk}^2$ for the shear rate in the cases $\alpha=0.9$ and $\alpha=0.8$ are marked with dotted-dashed vertical lines.}}
\label{phi}
\end{center}\end{figure}

In figure \ref{eta} we present the results for the shear viscosity $\eta_1^*$. {As we can observe, the kinetic model predicts that $\eta_1^*$ is practically independent of the mass ratio $\mu$. This feature is also shared by the DSMC data, except near $a_\kk$, where the values of $\eta_1^*$ increase with the mass ratio. Apart from this, the kinetic model not only successfully captures the decrease of the shear viscosity  with increasing shear rate (shear thinning), but also exhibits a generally good quantitative agreement.}

The normal stress coefficients $\theta_{1,x}$ and $\theta_{1,y}$ are displayed in figures \ref{thetax} and \ref{thetay}, respectively. While $\theta_{1,x}>1$, we observe that $\theta_{1,y}<1$, this anisotropic effect increasing with increasing shear rate, with increasing mass ratio, and with increasing collisional dissipation. The kinetic model correctly accounts for these trends. At a quantitative level, the agreement is fairly good for $\mu\leq 1$ (especially in the case of $\theta_{1,x}$). However, in the case of heavy impurities ($\mu>1$), {the kinetic model  clearly underestimates the deviations of $\theta_{1,x}$ and $\theta_{1,y}$ from unity. }

{The heat flux transport coefficients $\lambda_1^*$ and $\phi_1^*$ are shown in figures \ref{lambda} and \ref{phi}, respectively. The theoretical curves for $\lambda_1^*$ and, especially, for $\phi_1^*$ are more sensitive to the value of the mass ratio $\mu$ than those for $\eta_1^*$, showing that the heat flux transport coefficients increase with increasing $\mu$. These features are generally confirmed by the DSMC data for $\phi_1^*$ but in the case of  $\lambda_1^*$ the influence of $\mu$ is less clear, except near the threshold shear rate. Apart from that, the kinetic model correctly describes the decrease of $\lambda_1^*$ with increasing shear rate as well as the change in the dependence of $\phi_1^*$ on $a$ as one goes from the elastic case ($\alpha=1$) to the inelastic ones ($\alpha=0.9$ and $0.8$). It is interesting to remark that the cross thermal conductivity coefficient $\phi_1^*$ (an obvious non-Newtonian effect) can become larger than the generalized NS thermal conductivity coefficient $\lambda_1^*$. This effect was already observed in the special case of LTu flows in one-component systems \cite{VSG10,VGS11}. Now, both the kinetic model and the DSMC data show that the inequality $\phi_1^*>\lambda_1^*$ (i.e., $|q_{1,x}|>|q_{1,y}|$) becomes more pronounced as the shear rate or the mass ratio increase.}

\section{Conclusions\label{sec4}}
We have analyzed in this paper the properties of the Couette flow for a granular impurity {immersed in a granular gas}. We have focused on the region of high shear rates, i.e., $a>a_{\kk}$, {where viscous heating prevails over inelastic cooling and thus the thermal curvature coefficient $\gamma$ defined by equation \eref{III.7} is positive}.
This corresponds to states in the region above the LTu surface described in a previous work \cite{VSG10,VGS11},  where  an analytical solution from a BGK-type model is available \cite{VGS08} and, more importantly, nonlinear effects are dominant.  We have confirmed that the hypotheses made in order to obtain the BGK-type theoretical solution in a previous work \cite{VGS08,TTMGSD01} also apply for the numerical solution of the inelastic Boltzmann and Boltzmann--Lorentz equations:
{(i) the pressure $p_2$ of the granular gas is uniform, (ii) the local shear rate $\partial u_{2,x}/\partial y$ divided by the local collision frequency $\nu_2\propto n_2 T_2^{1/2}$ is uniform, (iii) the temperature $T_2$ of the granular gas is a quadratic function of the  flow velocity $u_{2,x}$, (iv) the flow velocity  of the impurity coincides with that of the gas, i.e., $u_{1,x}=u_{2,x}$, (v) the mole fraction $n_1/n_2$ of the impurity is uniform, and (vi) the impurity/gas temperature ratio $\chi\equiv T_1/T_2$ is also uniform.}  The fourth  hypothesis on the absence of mutual diffusion is fulfilled with an especially high degree of accuracy, since we were unable to measure any non-negligible difference between the profiles $\mathbf{u}_{1,x}(y)$ and {$\mathbf{u}_{2,x}(y)$} over a wide range of situations [see figure \ref{gr3p2}(a) for reference]. This is important since it delimits well the properties of the Couette flow for the impurity, which has important consequences in applications such as segregation \cite{GV10}.

{Apart from the validation of the hydrodynamic hypotheses (i)--(vi), the DSMC results have shown a good semi-quantitative validity of the theoretical predictions as regards the dependence  of non-Newtonian properties (temperature ratio, normal stress differences,  generalized shear viscosity, and  generalized thermal conductivities) on the shear rate, the mass ratio, and the coefficient of restitution.}

Although we have limited our study to the region $a>a_{\kk}$ (i.e., $\gamma>0$), preliminary results seem to indicate that the special LTu state ($a=a_\kk$ or $\gamma=0$) also exists for the granular impurity (and exactly at the same point that occurs for the granular gas). This, together with the fact that the impurity in any case fulfills also the hypotheses (i)--(iii) for the granular gas, suggests that we can perform a classification of Couette flows for the impurity analogous to that for the granular gas \cite{VU09,VSG10}. Therefore, an interesting way in which our work can be extended consists of studying  Couette flows for non-equal temperature walls and in the region below $a_{\kk}$. Theoretical and numerical work  on this subject is ongoing.

\ack
This research  has been supported by the Ministerio de  Ciencia e Innovaci\'on (Spain) through Grants No.\ {FIS2010-16587} and (only FVR)  No.\ {MAT2009-14351-C02-02}, partially financed by FEDER funds. Support from the Junta de Extremadura (Spain) through Grant No.\ GR10158 is also gratefully acknowledged.

\appendix
\section{Transport properties of the impurity from the BGK-type kinetic model
\label{appA}}

We present in this appendix the explicit expressions for the relevant  hydrodynamic properties, as extracted from a previous work \cite{VGS08}. Those expressions are given in terms of some mathematical functions that we define below.

First, we introduce the functions
\beq
\label{III.12} F_{0,m}(y,z)\equiv \int_{0}^{\infty} \dd w\ e^{-(1+z)w} w^m X_0\left(\Theta(w,y,z)\right),
\eeq
\beqa
\label{0.7}
F_{1,m}(y,z)&\equiv& y\frac{\partial}{\partial y} F_{0,m}(y,z)\nn
&=&-\frac{1}{2}\int_{0}^{\infty} \dd w\ e^{-(1+z)w} w^m \frac{X_1\left(\Theta(w,y,z)\right)}{
\Theta(w,y,z)}
\eeqa
where
\beq
\fl
X_0(\Theta)\equiv \sqrt{\pi}\Theta e^{\Theta^2}\text{erfc}\left(\Theta\right)-1,\qquad X_1(\Theta)\equiv \Theta^2\left[\sqrt{\pi}(1+2\Theta^2)e^{\Theta^2}\text{erfc}\left(\Theta\right)-2\Theta\right],
\label{X}
\eeq
\begin{equation}
\label{III.14}
\Theta(w,y,z)\equiv\frac{1}{2\sqrt{2y}} \frac{z}{1- e^{-\frac{1}{2}z w}}.
\end{equation}
In equation  \eref{X}, $\text{erfc}(x)$ is the complementary error function.
Note that, because of the square root in equation \eref{III.14}, the functions $F_{0,m}(y,z)$ and $F_{1,m}(y,z)$ are only well defined for $y\geq 0$.
Next, we define
\begin{equation}
\label{III.13}
G(y,z)\equiv \int_{0}^{\infty} dw\, e^{-(1+\frac{3}{2}z)w} w
\left[\frac{d+1}{2}X_1(\Theta(w,y,z))+Y(\Theta(w,y,z))\right],
\eeq
\begin{equation}
\label{III.13bis}
H(y,z)\equiv \int_{0}^{\infty} dw\, e^{-(1+\frac{3}{2}z)w} w^3
Y(\Theta(w,y,z)),
\eeq
where
\beq
Y(\Theta)\equiv\Theta^3\left[2(1+\Theta^2)-\sqrt{\pi}\Theta(3+2\Theta^2)e^{\Theta^2}\text{erfc}\left(\Theta\right)\right].
\label{Y}
\end{equation}

 Now that we have introduced the above functions, let us display the expressions for the solution of the kinetic model. First, the thermal curvature coefficient $\gamma$ is given as a function of the reduced shear rate $a$ and the coefficient of restitution $\alpha_2$ through the implicit equation:
\beqa
\label{III.9}
\fl
d\frac{c_2\zeta_{2}^*}{1+c_2\zeta_{2}^*}-\frac{2c_2^2a^2}{(1+c_2\zeta_{2}^*)^3}&=&
2F_{1,0}(c_2^2\gamma,c_2\zeta_2^*)+dF_{0,0}(c_2^2\gamma,c_2\zeta_2^*)\nn
&&+c_2^2a^2\left[2F_{1,2}(c_2^2\gamma,c_2\zeta_2^*)+F_{0,2}(c_2^2\gamma,c_2\zeta_2^*)\right],
\eeqa
where
\beq
c_i\equiv \frac{2}{k_d(1+\alpha_i)}.
\eeq
Since the functions \eref{III.12} and \eref{0.7} are not defined for negative $y$, the representation (\ref{III.9}) exists only for $\gamma \geq 0$ or,
equivalently, for $a \geq a_\text{th}$, where the threshold value $a_\text{th}$ of the shear rate (corresponding to $\gamma=0$) is \cite{VGS08,TTMGSD01}
\begin{equation}
\label{III.15}
a_\text{th}^2=\frac{d}{2c_2}\zeta_{2}^*(1+c_2\zeta_{2}^*)^2.
\end{equation}
In the case $a=a_\text{th}$ the viscous heating is exactly balanced
by collisional cooling and the heat flux becomes uniform \cite{VSG10,VGS11,SGV09}.

Next, the temperature ratio $\chi=T_1/T_2$ is obtained from the implicit equation
\beqa
\label{n9}
d\left(\frac{T_1}{T_{12}}-\frac{1}{1+\zetat}\right)-\frac{2\widetilde{a}^2}
{(1+\zetat)^3}&=&2F_{1,0}(\widetilde{\gamma},\zetat)+dF_{0,0}(\widetilde{\gamma},\zetat)\nn
&&
+\widetilde{a}^2\left[2F_{1,2}(\widetilde{\gamma},\zetat)+F_{0,2}(\widetilde{\gamma},\zetat)\right].
\eeqa
Here,
\beq
\widetilde{a}\equiv \frac{\nu_2}{\nu_1}c_1 a,\qquad \widetilde{\gamma}\equiv \frac{\nu_2^2}{\nu_1^2}\frac{T_{12}}{T_1}\frac{\chi}{\mu}c_1^2\gamma,
\label{atilde}
\eeq
\beq
\zetat\equiv \frac{c_1\zeta_1}{\nu_1}=c_1\frac{d+2}{2d}\frac{\mu+\chi}{(1+\mu)^2\chi}(1-\alpha_{1}^2).
\label{III.19n}
\eeq
In equation \eref{III.19n} use has been made of equation \eref{zetasij} with $\mathbf{u}_1=\mathbf{u}_2$. Moreover, from equations \eref{nu2}, \eref{nu1}, and \eref{4.18}, we have
\begin{equation}
\label{III.18n}
\frac{\nu_{2}}{\nu_1}=
\left(\frac{2}{1+\omega}\right)^{d-1}\sqrt{\frac{2\mu}{\mu+\chi}},\qquad \frac{T_{12}}{T_1}=1+\frac{2\mu(1-\chi)}{(1+\mu)^2\chi}.
\end{equation}
We recall that $\omega\equiv\sigma_1/\sigma_2$.

Finally, the transport coefficients defined by equations \eref{Pxy}--\eref{qx} are
\begin{equation}
\label{III.29}
{\eta_1^*=c_1\frac{T_{12}}{T_1}\left[\frac{1}{(1+\zetat)^2}+F_{0,1}(\widetilde{\gamma},\zetat)+
2F_{1,1}(\widetilde{\gamma},\zetat)\right],}
\end{equation}
\begin{equation}
\label{n7}
\theta_{1,x}=d-\frac{T_{12}}{T_1}\left[\frac{d-1}{1+\zetat}+(d-1)F_{0,0}(\widetilde{\gamma},\zetat)+2F_{1,0}(\widetilde{\gamma},\zetat)\right],
\end{equation}
\begin{equation}
\label{n6}
\theta_{1,y}=\frac{T_{12}}{T_1}\left[\frac{1}{1+\zetat}+F_{0,0}(\widetilde{\gamma},\zetat)+2F_{1,0}(\widetilde{\gamma},\zetat)\right]
,
\end{equation}
\beq
{\lambda_1^*=\frac{1}{d+2}\frac{T_{12}}{T_1}\frac{1}{\widetilde{\gamma}}\left[\eta_1^*\widetilde{a}^2-c_1\frac{d}{2}
\left(1-\frac{T_{12}}{T_1}+\zetat\right)\right],}
\label{III.32}
\eeq
\beq
{\phi_1=\frac{2c_1}{d+2}\left(\frac{T_{12}}{T_1}\right)^2\frac{\widetilde{a}}{\sqrt{2\widetilde{\gamma}}}
\left[G(\widetilde{\gamma},\zetat)+\widetilde{a}^2
H(\widetilde{\gamma},\zetat)\right],}
\label{III.33}
\eeq
where we have taken into account that $\text{Pr}=1$ in the BGK model \cite{C88}.

The transport coefficients for the granular gas are easily deduced from equations \eref{III.29}--\eref{III.33} by the formal replacements $1\to 2$, $\chi\to 1$, $\mu\to 1$, and $\omega\to 1$.
It is then easy to check that equation \eref{III.32} reduces to equation \eref{eta2}.

To conclude this appendix, let us write the results in the limit $a\to a_{\text{th}}$, i.e., $\gamma\to 0$. The temperature ratio is given by the physical root of the quartic equation \cite{VGS08}
\begin{equation}
\label{A6}
d\left(\frac{T_1}{T_{12}}-\frac{1}{1+\zetat}\right)-\frac{2\widetilde{a}_\kk^2}
{(1+\zetat)^3}=0,
\end{equation}
where $\widetilde{a}_\kk$ is obtained from equations \eref{III.15} and \eref{atilde}.
 Once
$\chi$ is known, the transport coefficients are \cite{VGS08}
\begin{equation}
\label{A7}
{\eta_1^*=c_1\frac{T_{12}}{T_1}\frac{1}{(1+\zetat)^2}},
\end{equation}
\begin{equation}
\label{A8}
\theta_{1,x}=d-(d-1)\theta_{1,y},\qquad \theta_{1,y}=\frac{T_{12}}{T_1}\frac{1}{1+\zetat} ,
\end{equation}
\beq
{\lambda_1^*=c_1\left(\frac{T_{12}}{T_1}\right)^2\frac{2}{2+7\zetat+6\zetat^2}
\left[1+\frac{6}{d+2}\frac{12+42\zetat+37\zetat^2}
{(2+7\zetat+6\zetat^2)^2}\widetilde{a}_\kk^2\right]},
\label{A9}
\eeq
\beq
\fl
{\phi_1^*=c_1\frac{2}{d+2}\left(\frac{T_{12}}{T_1}\right)^2\frac{4+7\zetat}{(2+7\zetat+6\zetat^2)^2}\widetilde{a}_\kk
\left[d+4+18\frac{8+28\zetat+25\zetat^2}
{(2+7\zetat+6\zetat^2)^2}\widetilde{a}_\kk^2\right].}
\label{A10}
\eeq
Note that equations \eref{III.15} and \eref{A7} (when the latter is particularized to the case of an impurity mechanically equivalent to the particles of the gas) are consistent with equation \eref{ath2}.

\section*{References}

\bibliographystyle{iopart-num}

\bibliography{D:/bib_files/Granular}

\providecommand{\newblock}{}
\begin{thebibliography}{10}
\expandafter\ifx\csname url\endcsname\relax
  \def\url#1{{\tt #1}}\fi
\expandafter\ifx\csname urlprefix\endcsname\relax\def\urlprefix{URL }\fi
\providecommand{\eprint}[2][]{\url{#2}}

\bibitem{AT06}
Aranson I~S and Tsimring L~S 2006 {\em Rev. Mod. Phys.\/} {\bf 78} 641--692

\bibitem{KJN93}
Knight J~B, Jaeger H~M and Nagel S~R 1993 {\em Phys. Rev. Lett.\/} {\bf 70}
  3728

\bibitem{JY02}
Jenkins J~T and Yoon D~K 2002 {\em Phys. Rev. Lett.\/} {\bf 88} 194301

\bibitem{TAH03}
Trujillo L, Alam M and Herrmann H~J 2003 {\em Europhys. Lett.\/} {\bf 64} 190

\bibitem{K04}
Kudrolli A 2004 {\em Rep. Prog. Phys.\/} {\bf 67} 209

\bibitem{BRM05}
Brey J~J, Ruiz-Montero M~J and Moreno F 2005 {\em Phys. Rev. Lett.\/} {\bf 95}
  098001

\bibitem{BRM06}
Brey J~J, Ruiz-Montero M~J and Moreno F 2006 {\em Phys. Rev. E\/} {\bf 73}
  031301

\bibitem{SGNT06}
Serero D, Goldhirsch I, Noskowicz S~H and Tan M~L 2006 {\em J. Fluid Mech.\/}
  {\bf 554} 237

\bibitem{ATH06}
Alam M, Trujillo L and Herrmann H~J 2006 {\em J. Stat. Phys.\/} {\bf 124} 587

\bibitem{G06a}
Garz\'o V 2006 {\em Europhys. Lett.\/} {\bf 75} 521

\bibitem{MPEU07}
Melby P, Prevost A, Egolf D~A and Urbach J~S 2007 {\em Phys. Rev. E\/} {\bf 76}
  051307

\bibitem{G08}
Garz\'o V 2008 {\em Phys. Rev. E\/} {\bf 78} 020301(R)

\bibitem{G09}
Garz\'o V 2009 {\em Eur. Phys. J. E\/} {\bf 29} 261

\bibitem{SNTG09}
Serero D, Noskowicz S~H, Tan M~L and Goldhirsch I 2009 {\em Eur. Phys. J. Spec.
  Top.\/} {\bf 179} 221

\bibitem{GV10}
Garz\'o V and {Vega Reyes} F 2010 {\em J. Stat. Mech.\/}  P07024

\bibitem{K10}
Kudrolli A 2010 {\em Phys. Rev. Lett.\/} {\bf 104} 088001

\bibitem{BDKS98}
Brey J~J, Dufty J~W, Kim C~S and Santos A 1998 {\em Phys. Rev. E\/} {\bf 58}
  4638

\bibitem{CC70}
Chapman C and Cowling T~G 1970 {\em The Mathematical Theory of Non-Uniform
  Gases\/} 3rd ed (Cambridge University Press, Cambridge)

\bibitem{GS95}
Goldshtein A and Shapiro M 1995 {\em J. Fluid Mech.\/} {\bf 282} 75--114

\bibitem{GD02}
Garz\'o V and Dufty J~W 2002 {\em Phys. Fluids\/} {\bf 14} 1476–--1490

\bibitem{GVM09}
Garz\'o V, {Vega Reyes} F and Montanero J~M 2009 {\em J. Fluid Mech.\/} {\bf
  623} 387

\bibitem{JM87}
Jenkins J~T and Mancini F 1987 {\em J. Appl. Mech.\/} {\bf 54} 27

\bibitem{JM89}
Jenkins J~T and Mancini F 1989 {\em Phys. Fluids A\/} {\bf 1} 2050--2057

\bibitem{Z95}
Zamankhan P 1995 {\em Phys. Rev. E\/} {\bf 52} 4877

\bibitem{AW98}
Arnarson B and Willits J~T 1998 {\em Phys. Fluids\/} {\bf 10} 1324

\bibitem{WA99}
Willits J~T and Arnarson B 1999 {\em Phys. Fluids\/} {\bf 11} 3116

\bibitem{GDH07}
Garz\'o V, Dufty J~W and Hrenya C~M 2007 {\em Phys. Rev. E\/} {\bf 76} 031303

\bibitem{GHD07}
Garz\'o V, Hrenya C~M and Dufty J~W 2007 {\em Phys. Rev. E\/} {\bf 76} 031304

\bibitem{GV09}
Garz\'o V and Vega~Reyes F 2009 {\em Phys. Rev. E\/} {\bf 79} 041303

\bibitem{VU09}
{Vega Reyes} F and Urbach J~S 2009 {\em J. Fluid Mech.\/} {\bf 636} 279

\bibitem{C89}
Campbell C~S 1989 {\em J. Fluid Mech.\/} {\bf 203} 449--473

\bibitem{C90}
Campbell C~S 1990 {\em Annu. Rev. Fluid Mech.\/} {\bf 22} 57

\bibitem{G03}
Goldhirsch I 2003 {\em Annu. Rev. Fluid Mech.\/} {\bf 35} 267--293

\bibitem{SGD04}
Santos A, Garz\'o V and Dufty J~W 2004 {\em Phys. Rev. E\/} {\bf 69} 061303

\bibitem{VGS07}
{Vega Reyes} F, Garz\'o V and Santos A 2007 {\em Phys. Rev. E\/} {\bf 75}
  061306

\bibitem{VGS08}
{Vega Reyes} F, Garz\'o V and Santos A 2008 {\em J. Stat. Mech.\/}  P09003

\bibitem{GS96}
Goldhirsch I and Sela N 1996 {\em Phys. Rev. E\/} {\bf 54} 4458

\bibitem{SGN96}
Sela N, Goldhirsch I and Noskowicz S~H 1996 {\em Phys. Fluids\/} {\bf 8} 2337

\bibitem{SG98}
Sela N and Goldhirsch I 1998 {\em J. Fluid Mech.\/} {\bf 361} 41

\bibitem{L04}
Lutsko J~F 2004 {\em Phys. Rev. E\/} {\bf 70} 061101

\bibitem{VSG10}
{Vega Reyes} F, Santos A and Garz\'o V 2010 {\em Phys. Rev. Lett.\/} {\bf 104}
  028001

\bibitem{VGS11}
{Vega Reyes} F, Garz\'o V and Santos A 2011 {\em Phys. Rev. E\/} {\bf 83}
  021302

\bibitem{B94}
Bird G~I 1994 {\em Molecular Gas Dynamics and the Direct Simulation of Gas
  Flows\/} (Clarendon, Oxford)

\bibitem{DBS97}
Dufty J~W, Brey J~J and Santos A 1997 {\em Physica A\/} {\bf 240} 212--220

\bibitem{BP04}
Brilliantov N~V and P\"oschel T 2004 {\em Kinetic Theory of Granular Gases\/}
  (Oxford University Press, Oxford)

\bibitem{TTMGSD01}
Tij M, Tahiri E~E, Montanero J~M, Garz\'o V, Santos A and Dufty J~W 2001 {\em
  J. Stat. Phys.\/} {\bf 103} 1035--1068

\bibitem{B34}
Burnett D 1934 {\em Proc. London Math. Soc.\/} {\bf 40} 382

\bibitem{LR03}
Lockerby D~A and Reese J~M 2003 {\em J. Comp. Phys.\/} {\bf 188} 333--347

\bibitem{GS03}
Garz\'o V and Santos A 2003 {\em Kinetic Theory of Gases in Shear Flows.
  Nonlinear Transport\/} (Kluwer Academic Publishers, Dordrecht)

\bibitem{SGV09}
Santos A, Garz\'o V and {Vega Reyes} F 2009 {\em Eur. Phys. J. Spec. Top.\/}
  {\bf 179} 141

\bibitem{C88}
Cercignani C 1988 {\em {The Boltzmann Equation and Its Applications}\/} (New
  York: Springer--Verlag)

\bibitem{BDS99}
Brey J~J, Dufty J~W and Santos A 1999 {\em J. Stat. Phys.\/} {\bf 97} 281

\bibitem{F97}
Frezzotti A 1997 {\em Phys. Fluids\/} {\bf 9} 1329--1335

\bibitem{MS97}
Montanero J~M and Santos A 1997 {\em Phys. Fluids\/} {\bf 9} 2057

\bibitem{MGSB99}
Montanero J~M, Garz\'o V, Santos A and Brey J~J 1999 {\em J. Fluid Mech.\/}
  {\bf 389} 391--411

\bibitem{AS05}
Astillero A and Santos A 2005 {\em Phys. Rev. E\/} {\bf 72} 031309

\bibitem{VGS08b}
{Vega Reyes} F, Garz\'o V and Santos A 2008 Rheological properties of a
  granular impurity in the {Couette} flow {\em The XVth International Congress
  on Rheology\/} vol 1027 ed Co A, Leal G, Colby R and Giacomin A~J (Melville,
  NY: AIP Conference Proceedings) pp 953--955

\end{thebibliography}

\end{document}